\documentclass[letterpaper, english, reprint, aps, citeautoscript, groupedaddress, superscriptaddress, nofootinbib, longbibliography] {revtex4-1}

\usepackage[T1]{fontenc}
\usepackage[latin9]{inputenc}
\usepackage{babel}
\usepackage{textcomp}
\usepackage[Symbol]{upgreek}
\usepackage{calc}
\usepackage{amssymb}
\usepackage{amsmath}
\usepackage{graphicx}
\usepackage{adjustbox}
\usepackage{xfrac}
\usepackage{bm}
\usepackage{dsfont}
\usepackage[colorlinks=true,linkcolor=black, citecolor=blue, urlcolor=blue, unicode=true]{hyperref}


\begin{document}
\preprint{}

\title{Polarized inelastic neutron scattering of non-reciprocal spin waves in MnSi}

\newcommand{\ill}{Institut Laue-Langevin, 71 Avenue des Martyrs, CS 20156, 38042 Grenoble cedex 9, France}
\newcommand{\tum}{Physik-Department, Technische Universit\"at M\"unchen, James-Franck-Str. 1, 85748 Garching, Germany}
\newcommand{\mlz}{Heinz-Maier-Leibnitz-Zentrum (MLZ), Technische Universit\"at M\"unchen (TUM), Lichtenbergstr. 1, 85747 Garching, Germany}
\newcommand{\cologne}{Institut f\"ur Theoretische Physik, Universit\"at zu K\"oln, Z\"ulpicher Str. 77a, 50937 K\"oln, Germany}
\newcommand{\dresden}{Institut f\"ur Theoretische Physik, Technische Universit\"at Dresden, 01062 Dresden, Germany}
\newcommand{\karlsruhe}{Institut f\"ur Theoretische Festk\"orperphysik, Karlsruhe Institute of Technology, 76131 Karlsruhe, Germany}
\newcommand{\karlsruhefkp}{Institut f\"ur Festk\"orperphysik, Karlsruhe Institute of Technology, 76344 Eggenstein-Leopoldshafen, Germany}

\author{T. Weber}
\email[Corresponding author: Tobias Weber, ]{tweber@ill.fr}
\affiliation{\ill}

\author{J. Waizner}
\affiliation{\cologne}

\author{P. Steffens}
\affiliation{\ill}

\author{A. Bauer}
\affiliation{\tum}

\author{C. Pfleiderer}
\affiliation{\tum}

\author{M. Garst}
\affiliation{\dresden}
\affiliation{\karlsruhe}
\affiliation{\karlsruhefkp}

\author{P. B\"oni}
\affiliation{\tum}

\date{\today}

\begin{abstract}
We report spin-polarized inelastic neutron scattering of the dynamical structure factor of the conical magnetic helix in the cubic chiral magnet MnSi. We find that the spectral weight of spin-flip scattering processes is concentrated on single branches for wavevector transfer parallel to the helix axis as inferred from well-defined peaks in the neutron spectra. In contrast, for wavevector transfers perpendicular to the helix the spectral weight is distributed among different branches of the magnon band structure as reflected in broader features of the spectra. Taking into account the effects of instrumental resolution, our experimental results are in excellent quantitative agreement with parameter-free theoretical predictions. Whereas the dispersion of the spin waves in MnSi appears to be approximately reciprocal at low energies and small applied fields, the associated spin-resolved spectral weight displays a pronounced non-reciprocity that implies a distinct non-reciprocal response in the limit of vanishing uniform magnetization at zero magnetic field.

\vspace{0.2cm}
\noindent This is a pre-print of our paper \cite{PublishedPaper} at \url{http://dx.doi.org/10.1103/PhysRevB.100.060404}, \\
\copyright{} 2018 American Physical Society.
\end{abstract}

\maketitle

The response of a condensed matter system is called non-reciprocal when it differs for stimuli with opposite wavevectors ${\bf q}$ and $-{\bf q}$. Prominent examples are the magnetochiral dichroism as well as the non-reciprocal magnon transport in systems with broken time-reversal symmetry \cite{Tokura2018}. In these systems, the time-reversal symmetry is often broken by a uniform external or internal magnetic field. Thin ferromagnetic films, for example, support Damon-Eshbach surface spin waves \cite{Camley1987} whose propagation is strongly non-reciprocal, which allows for unidirectional heat transport \cite{An:2013}. Non-reciprocal responses in the absence of uniform magnetic fields have been addressed in carefully tailored systems \cite{Otalora2016}.

\begin{figure}[t]
\includegraphics[width=0.4\textwidth]{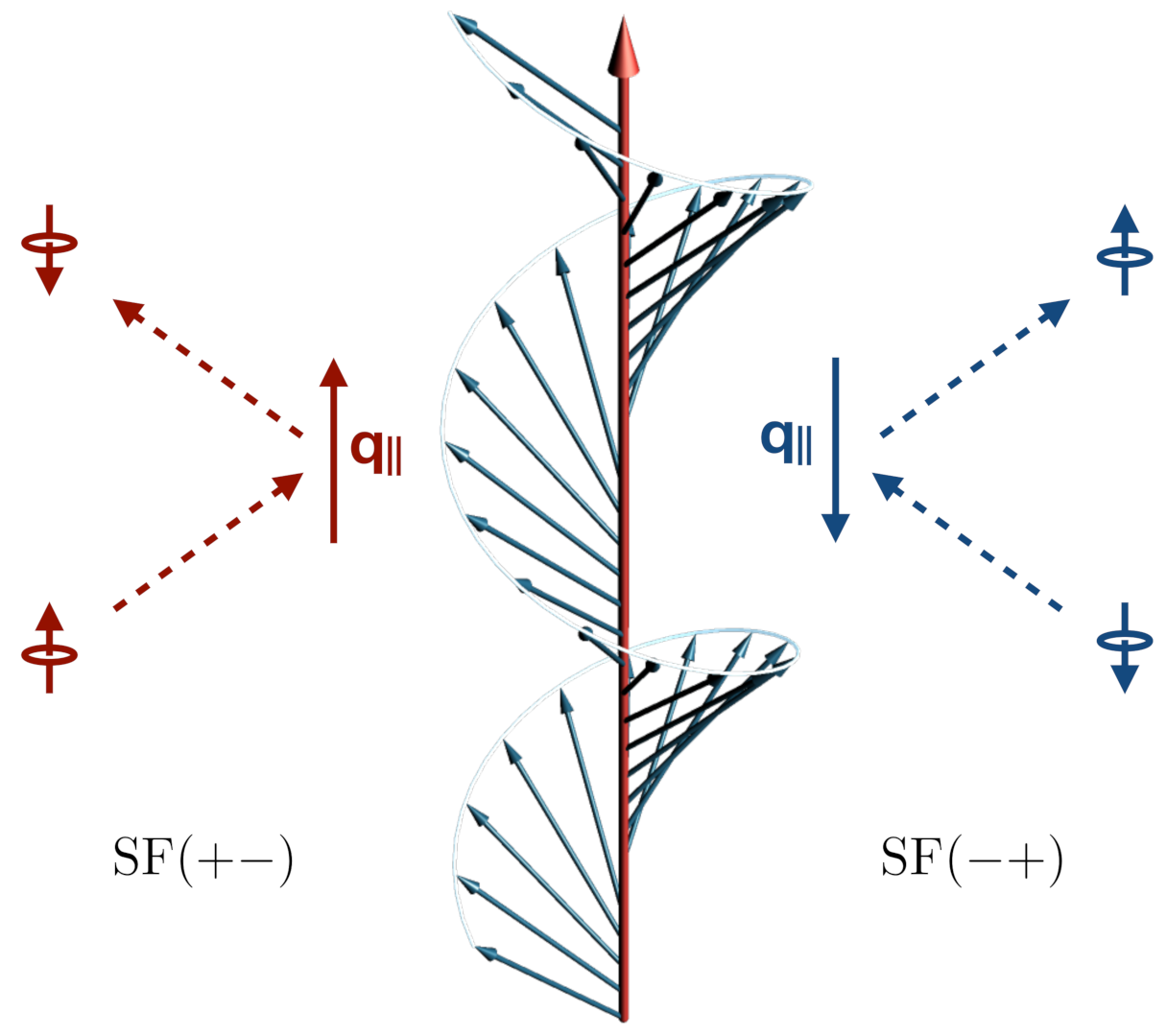}
\caption{Inelastic scattering processes between a neutron and a left-handed conical magnetic helix in the two spin-flip scattering channels SF($+-$) and SF($-+$). During the spin-flip process angular momentum $\pm \hbar$ is transferred to the helix exerting a torque. If the additional wavevector transfer is along the helix axis and commensurate, ${\bf q}_\parallel \approx \pm {\bf k}_h$, these processes activate a Goldstone spin wave mode associated with the broken translation symmetry. This figure was adapted from Ref.~\onlinecite{Garst2017} under CC BY 3.0 \cite{CCBY30}.}
\label{fig:archimedean}
\end{figure}

In this paper we focus, in contrast, on magnetic bulk materials with broken inversion symmetry. The non-centrosymmetric transition metal compound MnSi with its non-centrosymmetric cubic B20 crystal structure exhibits homo-chiral helimagnetic order due to the Dzyaloshinskii-Moriya spin-orbit interaction \cite{BauerPfleiderer2016}. The helimagnet spontaneously breaks time-reversal symmetry but without producing a uniform internal field. The magnetization winds around the helix axis in such a manner that the magnetization, in the absence of an external field, vanishes on average. For finite fields, however, the magnetic moments cant towards the field giving rise to a conical arrangement, see Fig.~\ref{fig:archimedean}.

Previous neutron scattering studies provided basic evidence for non-reciprocal spin waves in MnSi. Sato {\it et al.} \cite{Sato16} and Grigoriev {\it et al.} \cite{Grigoriev15} investigated the dynamics of spin waves in the field-polarized state where a large applied field fully polarizes the magnetic moments. In Refs.~\onlinecite{Jano10, Kugler15,Weber2017Field} the magnetic excitations at zero and finite fields were studied. However, either multiple magnetic domains or the scattering geometry did not allow to disentangle contributions from various scattering channels unambiguously. Moreover, these previous works were carried out by means of unpolarized neutrons forbidding a quantitative analysis of the non-reciprocal response encoded in the dynamical structure factor. While conceptually important, these studies therefore did not permit to draw any conclusions whether the helimagnon excitations in MnSi at small applied field maintain their non-reciprocal characteristics as predicted theoretically \cite{Weber2017Field}.

To address this issue, we report in this Letter spin-polarized inelastic neutron scattering under a carefully designed scattering geometry in order to resolve the contributions from both spin-flip scattering channels.  Our experimental data are in excellent quantitative agreement with the same theoretical framework that also accounted for the previous experiments. Whereas the application of a small magnetic field was necessary in order to prepare the sample in a single domain state of helimagnetic order, the field was sufficiently small so that we can extrapolate our findings with confidence to zero field where the induced uniform magnetic field vanishes. As our main results, we provide evidence for the non-reciprocal response of helimagnetic order in the structure factor, $\mathcal{S}_{\pm\mp}({\bf q},E) \neq \mathcal{S}_{\pm\mp}(-{\bf q},E)$, of both spin-flip scattering channels, where we define non-reciprocity in accordance with the definitions in Ref.~\onlinecite{Tokura2018}.

Non-reciprocal magnon transport is usually associated with an asymmetric dispersion $\varepsilon({\bf q}) \neq \varepsilon(-\bf{q})$ for the spin waves. This arises, in particular, in the field-polarized state of chiral magnets \cite{Melcher1973,Kataoka1987}. As a consequence, the group velocity $\partial_{\bf q} \varepsilon({\bf q})$ remains finite in the limit of small wavevector, ${\bf q} \to 0$, which is routinely used to determine the size of the Dzyaloshinskii-Moriya interaction \cite{Zakeri2010,Di2015,Belmeguenai2015,Seki:2016,Takagi2017}.
The spin waves maintain a non-reciprocal character also in the other long-range ordered phases. In particular, in the skyrmion lattice phase, that forms in cubic chiral magnets at intermediate fields close to the critical temperature \cite{Muehl09}, the spin wave dispersion is non-reciprocal for wavevectors $\bf{q}$ along the skyrmion strings \cite{Xing19,Lin2019,Kravchuk2019}, which was recently demonstrated with spin wave spectroscopy \cite{Seki2019}.

\begin{figure}[t]
\includegraphics[width=0.5\textwidth]{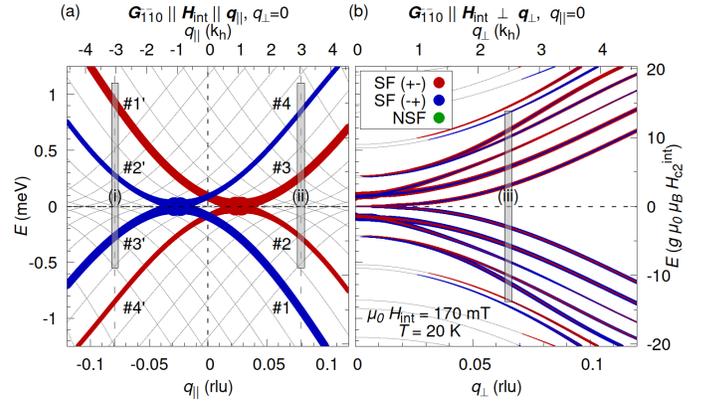}
\caption{Low-energy spin wave dispersion and spectral weight for the conical helix according to the theory outlined in Ref.~\onlinecite{Weber2017Field}. The axes label wavevectors and energies both in dimensionless and dimensionfull units where the latter was chosen for MnSi at 20 K with the reciprocal pitch $k_h = 0.036$ \AA$^{-1}$, lattice constant $a = 4.558$\AA, the critical internal field $\mu_0 H^{\rm int}_{c2} \approx 530$ mT, and $g \approx 2$ so that $g\mu_B \mu_0 H^{\rm int}_{c2} \approx 0.06$ meV. Panel (a) and (b) show, respectively, the dependence on the wavevector longitudinal, ${\bf q}_\parallel$, and transversal, $q_\perp = \left|{\bf q}_\perp\right|$, to the magnetic field with $\mu_0 H_{\rm int} = 170$ mT. The thin gray lines represent the dispersion whereas the width of the red and blue lines indicate the spectral weight of $\chi''_{\pm\mp}({\bf q},E)$ in the two spin-flip scattering channels SF($\pm\mp$), where positive and negative energies correspond, respectively, to the creation and annihilation of a magnon. Close to the helimagnetic satellites, $q_\parallel = \pm k_h$, we depict the weights in a logarithmic scale due to their strong increase in that region. Note that in the configuration with the nuclear reciprocal lattice vector ${\bf G}_{\bar 1 \bar 10} \parallel {\bf H}_{\rm int}$ non-spin-flip scattering (NSF) does not contribute. Experimental energy scans were performed within the gray shaded areas labeled $(i)-(iii)$, see Figs.~\ref{fig:exp-parallel} and \ref{fig:exp-perp}.}
\label{fig:theo}
\end{figure}

The evolution of non-reciprocity from the field-polarized, $H > H_{c2}$, to the conical helix phase, $H < H_{c2}$, was discussed in our previous work \cite{Weber2017Field}. The periodicity of the conical helix in general gives rise to Bragg scattering of spin waves resulting in a magnon band structure \cite{Kataoka1987,Jano10, Kugler15,Garst2017}. It is periodic for the wavevector component ${\bf q}_\parallel$ that is parallel to the helix wavevector, ${\bf k}_h \parallel {\bf H}$. The corresponding first Brillouin zone extends from $-k_h/2$ to $k_h/2$ where $k_h = |{\bf k}_h|$. Due to this backfolding and the symmetries of the conical helix, the dispersion becomes approximately symmetric $\varepsilon({\bf q}_\parallel) \approx \varepsilon(-\bf{q}_\parallel)$ at low energies, see the thin gray lines in Fig.~\ref{fig:theo}. The small deviations of this reflection symmetry in Fig.~\ref{fig:theo}(a) are attributed to corrections to the low-energy theory that are negligible for small energies $E$ \cite{Weber2017Field}.
However, the non-reciprocity still remains pronounced for $H < H_{c2}$ in the dynamical structure factor $\mathcal{S}_{ij}({\bf q}, E) = 2 (1+n_B(E)) \chi''_{ij}({\bf q}, E)$, where $n_B$ is the Bose function.
For example, the imaginary component of the susceptibility $\chi''_{+-}({\bf q}, E)$ describes the probability to scatter a neutron with transfer of energy $E$ and wavevector ${\bf q}$ in the spin-flip scattering channel SF($+-$), i.e., the incoming neutron possesses a spin $\uparrow$ and the outgoing neutron leaves with a spin $\downarrow$ transferring a total angular momentum of $\hbar$, i.e., reduced Planck constant, to the magnetic system.
A non-reciprocal response implies that $\chi''_{ij}({\bf q}, E) \neq \chi''_{ij}(-{\bf q}, E)$ \cite{Tokura2018}, whereas the relation $\chi''_{ij}({\bf q}, E) = - \chi''_{ji}(-{\bf q}, - E)$ always holds by definition.

Spin-flip scattering processes where the neutron transfers a momentum that is parallel to the helix wavevector, ${\bf q} = {\bf q}_\parallel$, see Fig.~\ref{fig:archimedean}, are particularly interesting due to the screw symmetry of the helix. The orientation of, e.g., a left-handed magnetic helix for a field along the $z$-axis is described by the unit vector $\hat n(z) = \sin \theta (\cos (k_h z), -\sin(k_h z),0) + \cos\theta (0,0,1)$ with the cone angle $\theta$. Its screw symmetry can be mathematically expressed by
\begin{align}
(\hat P_z - k_h \hat L_z) \hat n = 0
\end{align}
where $\hat P_z = - i\mathds{1} \hbar \partial_z$ is the momentum operator along the $z$-axis defined by the field and $(\hat L_z)_{ij} = - i \hbar \varepsilon_{ijz}$ is the $z$-component of a spin-1 operator generating rotations around the $z$-axis, where $\varepsilon_{ij\ell}$ is the totally antisymmetric tensor. In case the  momentum and angular momentum transferred in the scattering processes are consistent with this screw symmetry, $\hbar {\bf q}_\parallel \approx \pm \hbar {\bf k}_h$ and $\pm \hbar$, respectively, a Goldstone spin wave mode with vanishing energy is activated with a large spectral weight close to $E=0$.

The screw symmetry also ensures that for wavevectors along the helix axis, ${\bf q} = {\bf q}_\parallel$, the spectral weight of each Goldstone mode is concentrated on a single branch in the magnon band structure. Considering a single spin-flip channel, the resulting scattering probability is thus strongly non-reciprocal as a function of ${\bf q}_\parallel$, see Fig.~\ref{fig:theo}(a), and it remains so even at zero magnetic field.

As a function of a wavevector ${\bf q}_\perp$ perpendicular to the helix axis, the spectral weight gets instead distributed among different branches of the  band structure, see Fig.~\ref{fig:theo}(b). Neglecting magnetocrystalline anisotropies, the spectrum as well as the spectral weight only depend on the magnitude of the perpendicular component $q_\perp = |{\bf q}_\perp|$ \cite{Weber2017Field}. For a finite magnetic field the scattering probability is still non-reciprocal in Fig.~\ref{fig:theo}(b) with respect to inversion of the energy transfer, $E \to -E$. The probability to create a magnon, $E>0$, is larger in the SF($-+$) channel, whereas the probability to absorb a magnon, $E<0$, is larger in the SF($+-$) channel.

In this paper we report inelastic scattering experiments with linearly polarized neutrons using the cold-neutron triple-axis spectrometer ThALES \cite{thales} at the Institut Laue-Langevin in Grenoble, France. For our study ThALES was set up for longitudinal polarization analysis using the $(111)$ reflection of $\mathrm{Cu_2MnAl}$ Heusler crystals permitting both polarization and analysis of the neutrons \cite{Moon69}. Spin flipper coils were placed before and after the sample. Higher order neutrons in the incident beam were removed by a neutron velocity selector. A single collimator allowing a maximum horizontal beam divergence of 30 minutes was installed between the sample and the analyzer position. Neutron energies of $E_{f,1} = 3.5$ meV and $E_{f,2} = 4.06$ meV were used. At the $(111)$ reflection they correspond to calculated incoherent (coherent) instrumental resolutions of $\Delta E_{f,1}^{\mathrm{inc}} = 0.055\ \mathrm{meV}$ ($\Delta E_{f,1}^{\mathrm{coh}} = 0.049\ \mathrm{meV}$) and $\Delta E_{f,2}^{\mathrm{inc}} = 0.072\ \mathrm{meV}$ ($\Delta E_{f,2}^{\mathrm{coh}} = 0.06\ \mathrm{meV}$), respectively, where all values are given as full-width at half-maximum.

For the experiments, a cylindrical single-crystal of MnSi exhibiting left-handed helimagnetic ordering was used, which was already investigated in previous studies \cite{Bauer2010, Weber2017Field, Weber2018non}. It has a diameter and a height of 1 cm and 3 cm, respectively. With the cylinder axis vertical along the crystallographic $[001]$ direction, the set-up provided access to Bragg reflections in the $( hk0 )$ scattering plane. All measurements were performed at a temperature $T = 20$\,K around the $\bm{G}_{\bar 1 \bar 10}$ nuclear Bragg reflection in a magnetic field $\mu_0 H = 195\, \mathrm{mT}$ corresponding to an internal field $\mu_0 H_{\rm int} = 170\, \mathrm{mT}$ \cite{Sato89}. The pitch of the helix in MnSi at the temperature 20 K is $\lambda_ h \approx 175$ \AA\, corresponding to a helix wavenumber $k_h = 2\pi/\lambda_h = 0.036 $ \AA$^{-1}$. $\bm{H}$ was aligned along the $[110]$ direction, such that the sample was in a single-domain helical state with $\bm{k}_h \parallel [110]$.

Importantly, in this geometry the polarization axis of the neutrons, $\hat{P}$, was parallel to the applied field as well as to the nuclear reciprocal lattice vector, $\hat{P} \parallel {\bf H} \parallel \bm{G}_{\bar 1\bar 10}$.
For a total momentum transfer ${\bf Q} = \bm{G}_{\bar 1\bar 10} + {\bf q}$ with $|{\bf q}| \ll \bm{G}_{\bar 1\bar 10}$, only the two spin-flip scattering processes contribute to the magnetic scattering cross section. The non-splin-flip process does not contribute as it is orthogonal to the subspace of the projection operator $\mathds{1} - \hat Q \hat Q^T \approx \mathds{1} - \hat{P}\hat{P}^T$, where $\hat Q = {\bf Q}/|{\bf Q}|$, arising from the dipolar interaction between the neutron spin and the magnetization \cite{Moon69,Squires2012}.

\begin{figure}[t]
	\includegraphics[width=0.45\textwidth]{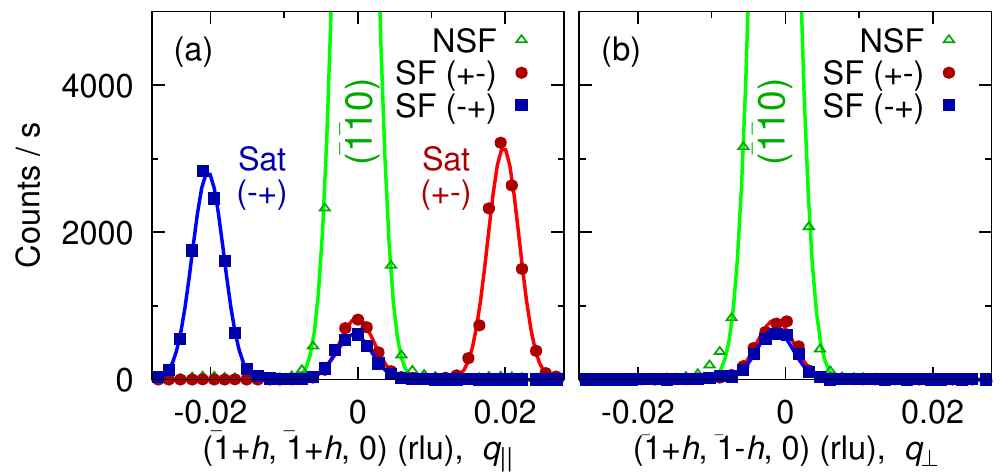}
	\caption{Elastic scans around the $(\bar 1\bar 10)$ Bragg peak of MnSi at $T = 20\,\mathrm{K}$ and $\mu_0 H_{\rm int} = 170\, \mathrm{mT}$. (a) Two magnetic satellites are observed at the positions $(-1.02, -1.02, 0)$ and $(-0.98,-0.98, 0)$ for spin-flip scattering SF$(-+)$ and SF$(+-)$, respectively, proving the single handedness of the crystal. The strong scattering at $(\bar 1\bar 10)$ is due to nuclear Bragg scattering. (b) The transverse scan indicates the excellent quality of the crystal.}
\label{fig:elast}
\end{figure}

Elastic scans with $\bm{q}$ longitudinal and transverse to $\bm{G}_{\bar 1\bar 10}$ were conducted. The $\bm{q_{\parallel}}$-scans shown in Fig.~\ref{fig:elast}(a) demonstrate that the two satellites belong indeed to different spin-flip channels confirming the single-handedness of the magnetic order. Nuclear scattering leads to the central $(\bar 1\bar 10)$ Bragg peak in the non-spin-flip channel. The small nuclear contribution at $(\bar 1\bar 10)$ in the nominal spin-flip channels may be attributed to the finite polarization of the neutron beam ($P = 91\pm1\%$). The $\bm{q_{\perp}}$-scans shown in Fig.~\ref{fig:elast}(b) indicate the high quality of the crystal, i.e. that the mosaic of the single crystal is small ($\eta_{\mathrm{FWHM}} = 18\, \mathrm{min}$) and well behaved.

\begin{figure}[t]
	{\includegraphics[width=0.4\textwidth]{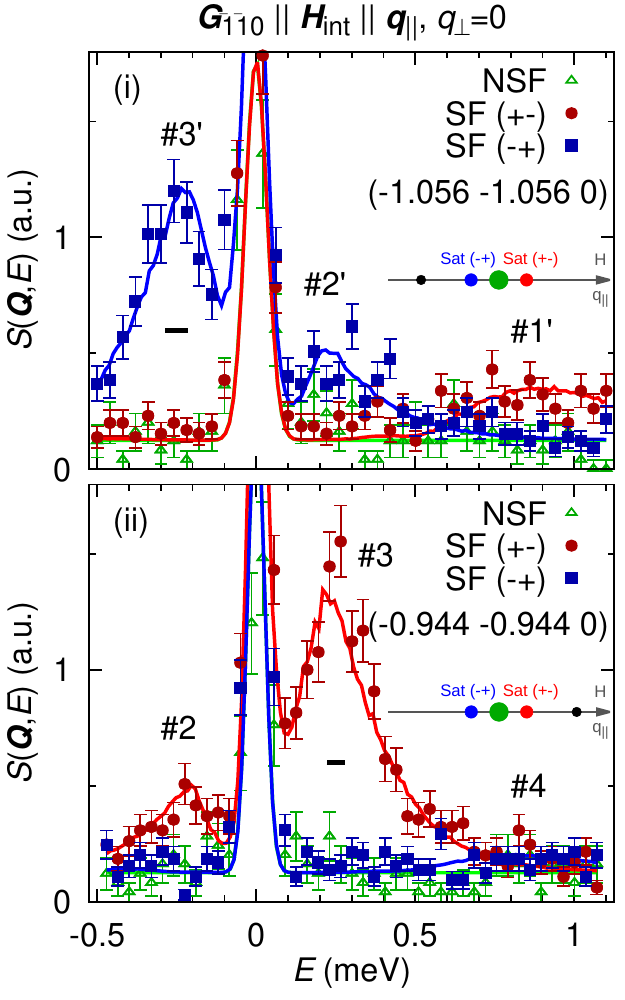}}
	\caption{Spin-polarized inelastic neutron scattering (symbols) with a wavevector transfer ${\bf q}_\parallel$ $(i)$ antiparallel and $(ii)$ parallel to the applied field. The energy scans correspond to the cuts $(i)$ and $(ii)$ indicated by the gray shaded areas in the theoretical spectrum of Fig.~\ref{fig:theo}(a). The observed peaks can be clearly assigned to the corresponding Goldstone spin wave modes. The solid lines are convolutions of the theoretical model with the instrumental resolution function. The scans were performed at $T = 20\,\mathrm{K}$ and $\mu_0 H_{\rm int} = 170\, \mathrm{mT}$.
	The thick black horizontal lines depict the calculated instrumental energy resolution. }
\label{fig:exp-parallel}
\end{figure}

\begin{figure}[t]
		{\includegraphics[width=0.4\textwidth]{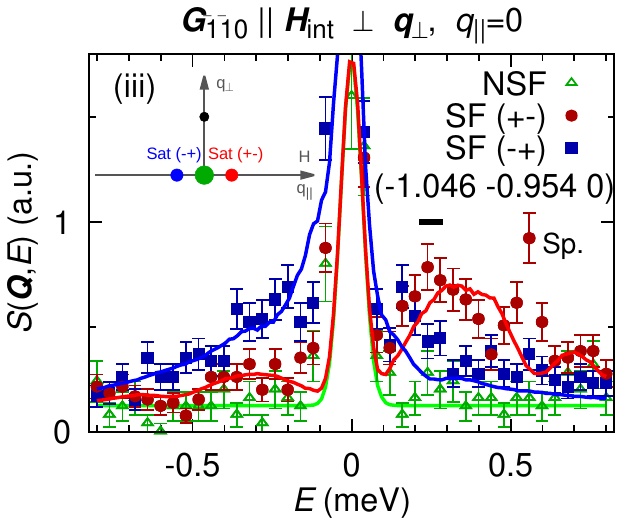}}
	\caption{Spin-polarized inelastic neutron scattering (symbols) for a wavevector transfer ${\bf q}_\perp$ perpendicular to the applied field at ${\bf q}_\parallel = 0$. The energy scans correspond to the cut indicated by the gray shaded area $(iii)$ in the calculated spectrum shown in Fig.~\ref{fig:theo}(b). The contributions of a large number of magnon bands result in broad features.  The solid lines are convolutions of the theoretical model with the instrumental resolution. Scans were recorded at $T = 20\,\mathrm{K}$ and $\mu_0 H_{\rm int} = 170\, \mathrm{mT}$. The data point marked \textit{``Sp.''} is a spurious instrumental artifact. The calculated instrumental energy resolution is shown as a thick black horizontal line. }
\label{fig:exp-perp}
\end{figure}

Shown in Fig.~\ref{fig:exp-parallel} are energy scans at the symmetric positions $\bm{Q_{(i)}} = (-1.056\ -1.056\ 0)\, \mathrm{rlu}$ and $\bm{Q_{(ii)}} =(-0.944\ -0.944\ 0)\, \mathrm{rlu}$ corresponding to a wavevector transfer to the magnetic system, ${\bf q} = {\bf q}_\parallel$, that is, respectively, antiparallel and parallel to the field.
The spectra are strongly non-reciprocal, i.e., the spectral weight at positive and negative energies are markedly different. In addition, the peak positions at positive and negative energies are approximately exchanged when comparing antiparallel and parallel wavevector transfer ${\bf q}_\parallel$ in panel $(i)$ and $(ii)$, respectively.
The experimental spectra for $\bm{Q_{(iii)}} = (-1.046\ -0.954\ 0)\, \mathrm{rlu}$ corresponding to a wavevector transfer ${\bf q}_\perp$ perpendicular to the field at ${\bf q}_\parallel = 0$ are shown in Fig.~\ref{fig:exp-perp}. The non-reciprocity with respect to the sign of the energy transfer $E$ is clearly visible. Broad excitations appear in both spin-flip channels denoted in red and blue but with very different intensities.

The experimental scans correspond to cuts in the theoretical spectra of Fig.~\ref{fig:theo} indicated by the gray shaded vertical bars. The peaks in Fig.~\ref{fig:exp-parallel} can be clearly assigned to the modes expected theoretically and we have labeled them accordingly. Due to the symmetry $\chi''_{+-}({\bf q},E) = - \chi''_{-+}(-{\bf q},-E)$, modes appear in pairs with the same spectral weight and we have denoted the partner at negative ${\bf q}_\parallel$ with a prime. For example, the modes \#3 and \#3' clearly dominate the spectra in Fig.~\ref{fig:exp-parallel}. The other peaks are weaker due to both their lower spectral weight and the smearing caused by the coarse resolution of the spectrometer at large $E$-transfers.
Note that the modes \#1 and \#4' are not observed because their energy could not be accessed in the experiment. A finite perpendicular wavevector ${\bf q}_\perp$ lowers the symmetry and activates Bragg scattering of spin waves due to the periodicity of the conical helix \cite{Kugler15}, which leads to a redistribution of spectral weight among different bands, see Fig.~\ref{fig:theo}(b). The contributions from the various bands cannot be resolved in the experiment giving rise to the broader features in the neutron spectra of Fig.~\ref{fig:exp-perp}.

For a quantitative comparison, we computed the dynamical structure factor $\mathcal{S}_{\pm\mp}(\bm{q}, E)$, which was convoluted with the resolution function \cite{Popovici1975} of ThALES using a full four-dimensional Monte-Carlo simulation \cite{TakinSource, Takin2017, Takin2016}. All theoretical parameters describing the magnetism of MnSi are known from previous experiments \cite{Weber2017Field, Weber2018non} and were kept fixed at their corresponding values. In order to adjust the overall intensity, a single relative normalization factor was simultaneously determined for all scans, similarly to the method of our preceding study \cite{Weber2017Field}. The resulting convolutions are shown as solid lines in Figs. \ref{fig:exp-parallel} and \ref{fig:exp-perp} where we find remarkably good agreement between theory and experiment.

In summary, using triple-axis spectroscopy with polarized neutrons we have confirmed and elucidated the non-reciprocal response associated with the conical helix in bulk samples of the cubic chiral magnet MnSi. Our results on the spontaneous properties of a bulk material provides an important point of comparison for the large body of work on tailored systems \cite{Otalora2016,Zakeri2010,Di2015,Belmeguenai2015}. Whereas the energy dispersion, $\varepsilon({\bf q}_\parallel)$, for wavevectors along the field ${\bf q}_\parallel$ is approximately reciprocal at low energies and small magnetic fields, necessary for the creation of a single domain state, the spin-polarized distribution of spectral weight remains non-reciprocal. The overall agreement with theoretical prediction implies a non-reciprocal response in the limit of zero magnetic field and thus vanishing uniform magnetization.

\begin{acknowledgements}
We acknowledge support by the DFG under Grant No. GE 971/5-1. M.G. is supported by the DFG through project A07 of SFB 1143 (project-id 247310070), GA 1072/5-1 and GA 1072/6-1. A.B. and C.P. are supported through DFG Grant No. TRR80 (project E1), SPP 2137 (Skyrmionics) and ERC Advanced Grants No. 291079 and 788031 (TOPFIT and ExQuiSid, respectively).
We are very grateful for the technical support by E. Villard and P. Chevalier, the additional instrumental support by M. B\"ohm as well as the instrumental control system support by J. Locatelli. We thank G. Brandl for providing his original implementation of the theoretical helimagnon model \cite{Kugler15} as reference. Fig.~\ref{fig:archimedean} was adapted from Ref.~\onlinecite{Garst2017} under CC BY 3.0 \cite{CCBY30}. The reported experiment has the DOI 10.5291/ILL-DATA.INTER-436 \cite{dataThales18}.
\end{acknowledgements}

\end{document}